В.С. Усатюк (аспирант)

# ОБЗОР СИСТЕМ АССИМЕТРИЧНОГО ШИФРОВАНИЯ НА ОСНОВЕ ЗАДАЧ ТЕОРИИ РЕШЕТОК КРИПТОСТОЙКИХ К КВАНТОВЫМ ВЫЧИСЛИТЕЛЬНЫМ МАШИНАМ

Братск, Братский государственный университет

Системы ассиметричного шифрования лежат в основе множества систем шифрования (PGP, S/MIME), сетевых протоколов (SSL, SSH, TLS и их производных), а также протоколах электронной цифровой подписи и в частности системах организации сетевой инфраструктуры (от сертификатов для удостоверения права на домен, реализующего привязку IP к доменному имени до транзакционных центров банков-экваеров).

Широко используемые на сегодняшний день ассиметричные системы шифрования основаны на двух типах задач теории чисел:
- задачах факторизации целых чисел (основаны на алгоритме RSA: схема Эль-Гамаля; криптосистема Рабина и прочие);
- задачах дискретного логарифмирования (семейства алгоритмов Диффи-Хелмана и их производные на других числовых полях: ECC - криптография на основе эллиптических кривых; ГОСТ 34.10-94; ГОСТ 34.10-2001).

Обращение этих задач, считалось неосуществимым за разумное время по причине отсутствия полиномиальных алгоритмов по времени выполнения. Начиная со второй половины 90-х годов 20 В. начинается период «квантовой революции» в теории алгоритмов.

В 1995 г. Питер Шор продемонстрировал полиномиальные алгоритмы обращения описанных выше задач на квантовых компьютерах [1], тем самым определив период существования перечисленных систем до возникновения квантовых вычислителей достаточной размерности. Эксперимент Чжуана в 2001 г. продемонстрировал выполнение алгоритма факторизации

Шора для числа 15 на 7-кубитном гибридном квантовом компьютере, построенного из $10^{18}$-молекул, состоящих из пяти атомов фтора и двух атомов углерода, с записью информации посредством радиосигналов и считыванием методами ядерного магнитного резонанса.

В 1996 г. Гровер продемонстрировал общий метод поиска в базе данных со сложностью $O(\sqrt{N})$, позволяющий реализовывать расшифровку симметричных алгоритмов шифрования эквивалентную двукратному уменьшению ключа шифра [2]. На практике работа алгоритма была проверена на 2-х кубитном квантовом компьютере, состоящем из полумиллитра смеси изотопа карбона-13 помеченного хлороформом, находящегося в ацетоне-D6[3].

В 1997 г. был представлен алгоритм Брассарда-Хойера-Таппа, основанный на алгоритме Гровера реализующий поиск коллизии хеш-функции со сложностью $O(\sqrt[3]{\frac{N}{r}})$, где N-мощность пространства перебора, а $r$ - число прообразов хеш-функции на один образ [4].

Таким образом, все типы используемых на практике ассиметричных систем шифрования, а так же ключевая стадия шифрования - хеширования переставили быть сложными для обращения. Появление этих результатов обусловило значительный рост интереса специалистов и финансирования данной тематике исследований, что позволило физикам и ряду утверждать возможность появления квантового компьютера необходимой размерности в ближайшие 15-20 лет. Возникла необходимость поиска криптостойких к квантовым компьютерам задач для постквантовой эпохи шифрования.

В 1996 г. венгерский математик-исследователь IBM Миклос Айтаи в своей работе [5] показал, что [6]:

– возможно, построить одностороннюю функцию на основе SVP-задачи, по базису решетки, найти кратчайший ненулевой вектор (shortest vector problem, SVP); более поздние исследователи улучшили результат до односторонней функции с секретом (trapdoor function) – вариантом од-носторонней функции, быстро обращаемой (по сравнению со скоростью получения образа функции) при наличии дополнительных сведений;
- переформулированная в вероятностный вариант задачи о рюкзаке, SVP-задача не имеет вероятностного полиномиального алгоритма решения, т.е. не разрешима за полиномиальное время на квантовых вычислителях;
- среди всего класса NP-задач, SVP-задача является самой сложной, т.е. является NP-полной задачей.

Результаты Айтая, а так же неудачные попытки реализации квантовых алгоритмов решения задач теории решеток, по аналогии с предложенным Шнором в 1987 г. эффективным мультиблочным редукционным дополнением [7] к полиномиальному алгоритму $L^3$ или LLL [8], Ленстры, Ленстры и Ловаса позволяющим приближенно решать SVP и сводимых к ним задачи с произвольно заданной точностью сделали эти задачи наиболее вероятными претендентами на реализацию криптостойких к квантовым машинам систем шифрования. Перед тем, как перейдем к дальнейшему изложению введем ряд ключевых понятий.

Решетка – дискретная аддитивная подгруппа, заданная на множестве $R^n$, т. е. решетку $L$ можно представить как множество векторов заданных целочисленными линейно независимыми базисными векторами $B = \{\overline{b_1},...,\overline{b_n}\} \subset R^n$, определенными по модулю некоторого целого числа $x \in Z^n$, $L = \sum_{i=1}^{n} \overline{b_i} \cdot Z = \{Bx : x \in Z^n\}$ (рис. 1). У решетки может быть множество базисов, $L = \sum_{i=1}^{n} \overline{a_i} \cdot Z$, (рис. 2). На рисунках 3, 4 показаны фундаментальные параллелепипеды образованные базисами. Площади (объемы в многомерном случае) фундаментальных параллелепипедов образованных всевозможными базисами одной решетки $L$, $\det(L)$ будут равны. Т.е. $\det L$ инвариант решетки. Под кратчайшим вектором решетки будем понимать вектор с координа-тами $\lambda_1(L) = \min_{x,y \in L, x \neq y} \|x - y\| = \min_{x \in L, x \neq 0} \|x\|$ (рис. 5). Тогда многомерным обобщением этого понятия будет $\lambda_i(L)$ - ограниченное минимальным r, для которого размерность решетки внутри шара радиуса r больше либо равна k (рис. 6).

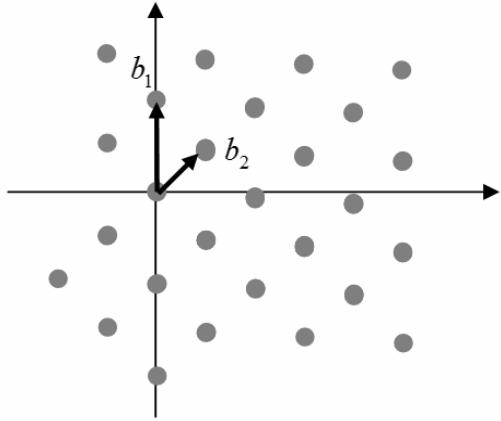 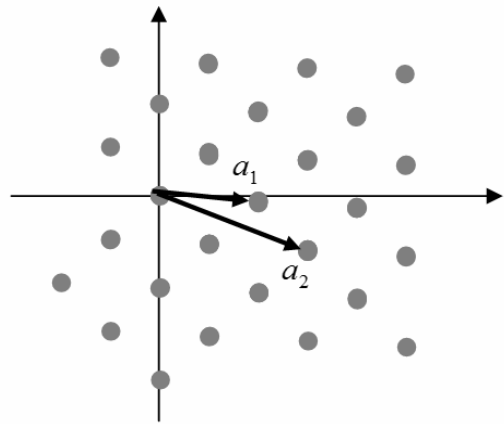

Рис. 1. Решетка с базисом $\{\overline{b}_1, \overline{b}_2\} \in B$   Рис. 2. Решетка с базисом $\{\overline{a}_1, \overline{a}_2\} \in B$

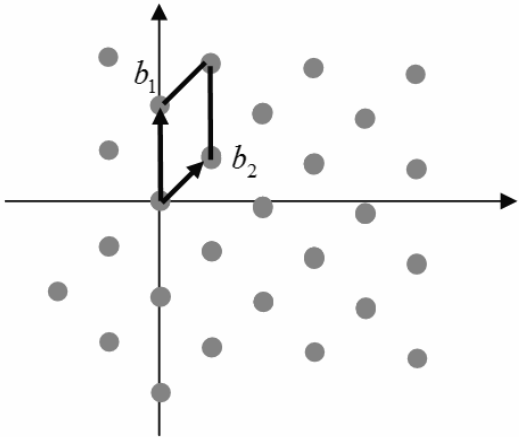 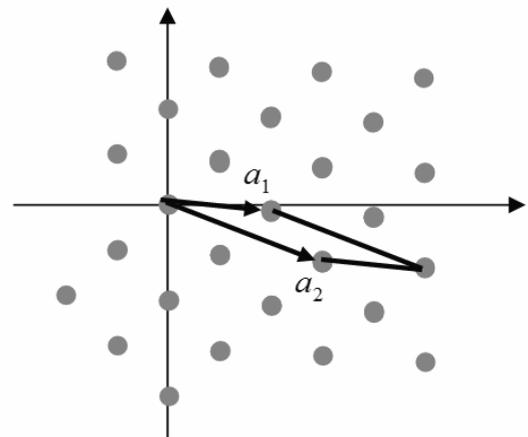

Рис. 3, 4. Фундаментальные параллелепипеды, образованные базисами

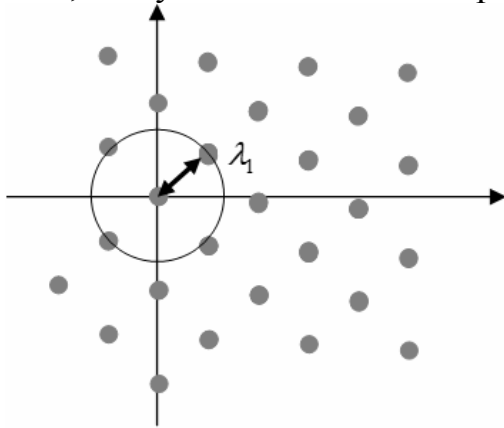 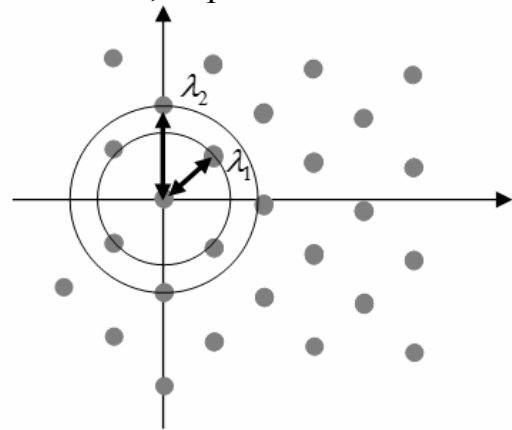

Рис. 5. Кратчайший вектор     Рис. 6. Кратчайший базис решетки $L$ в $R^2$

Таким образом, познакомившись с основными определениями теории решеток, перечислим задачи, применяемых при построении и оценке сложности в рассматриваемых нами системах шифрования:

1. По базису решетки, найти кратчайший ненулевой вектор (shortest vector problem, SVP), (рис. 7);
2. По базису решетки $L \in Z^{m \times n}$ и вещественному $\gamma > 0$, найти ненулевой вектор $\bar{b} \in LZ^n \setminus \{0\} : \|\bar{b}\|_p \leq \gamma \cdot \lambda^p_1(L)$ (shortest vector problem, $SVP^p_\gamma$), (рис. 8);

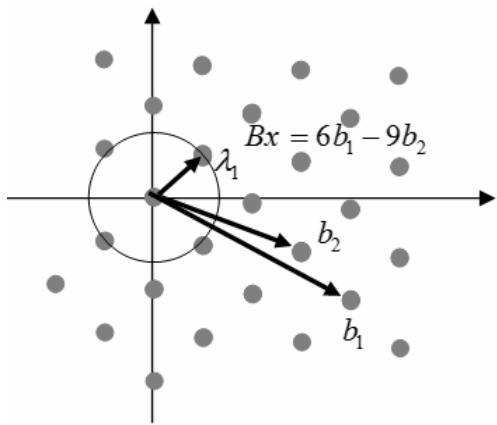 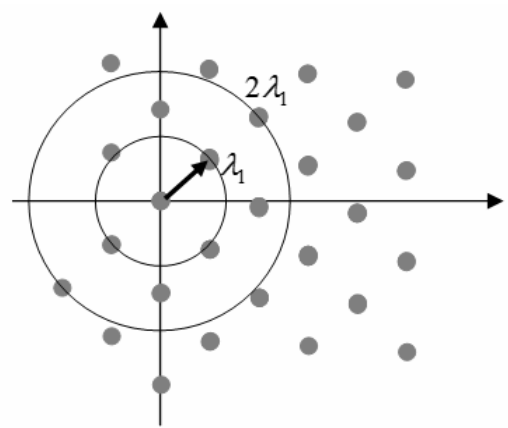

Рис. 7. Пример SVP-задачи в $R^2$     Рис. 8. Пример $SVP_\gamma$-задачи в $R^2$

3. По базису решетки, заданному вектору $\bar{j}$, найти ближайший вектор $\bar{b}$ к вектору $\bar{j}$ (Closes Vector Problem, CVP), (рис. 9);
4. По базису решетки $L \in Z^{m \times n}$, вещественному $\gamma > 0$ и заданному вектору $\bar{j} \in LR^n$, найти ненулевой вектор $\bar{b} \in LZ^n : \|\bar{j} - \bar{b}\|_p \leq \gamma \lambda^p_1(L)$ (Closes Vector Problem, $CVP^p_\gamma$);

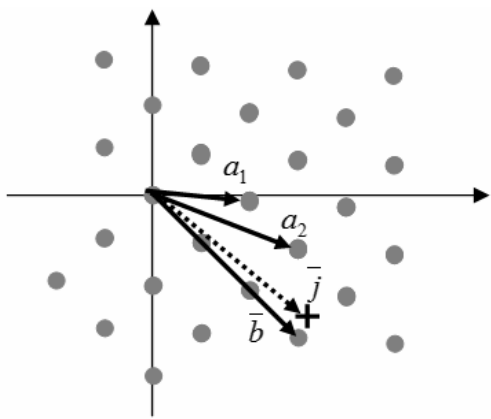 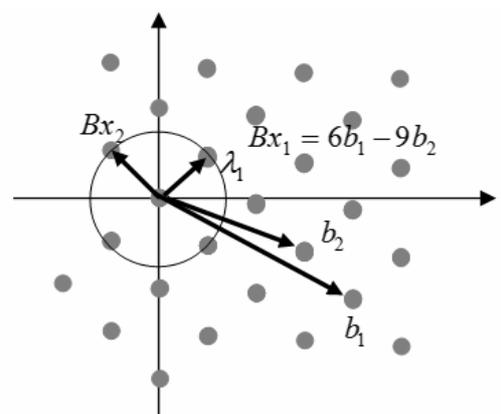

Рис. 9. Пример CVP-задачи в $R^2$     Рис. 10. Пример SIVP-задачи в $R^2$

5. Пусть дана n-мерная решетка L. Найти линейно независимые вектора $\bar{b}_1, ..., \bar{b}_n \in L$, для которых $\max_{i=1}^{n} \|\bar{b}_i\|_p \leq \gamma \lambda^p_n(L)$, где $\lambda^p_n(L)$, это длины n-соответствующих кратчайших векторов в решетке с p-нормой (Shortest Independent Vector Problem, $SIVP^p(n,\gamma)$), (рис. 10);

6. Пусть дана n-мерная решетка L. Найти вектор $\bar{u} \in L \setminus \{0\} : \|u\|_p \leq \gamma \lambda_1^p(L)$, где $\lambda_1^p(L)$, это длина кратчайшего вектора в решетке с p-нормой и $\bar{u}$ - кратчайший $\gamma$ - уникальный вектор, т.е. $\forall w \in L : \lambda_1^p(L) \leq \|\bar{w}\|_p \leq \gamma \lambda_1^p(L)$ $\bar{w} = z\bar{u}$ для некоторых $z \in Z$ (Unique Shortest Vector Problem, $uSVP^p(n,\gamma)$), (рис. 11);

7. Пусть дан базис B, q-нарной(модулярной) m-мерной решетки $L_q^{m \times n}$, т.е. решетка L, для которой принадлежность вектора к решетки L определяется: $L(B) = \{B^T s \bmod q \subseteq Z^m, s \in Z^n\}$. На решетке равномерно распределен шум $e$ (обычно с моментом ожидания равным 0 и дисперсией $\sqrt{q}$), q задан некоторым многочленом, $\bar{s} \in Z_q^n$ - некоторый исходный вектор без шума, известно значение $(B\bar{s} + \bar{e})$. Найти исходную точку в решетке (исключить шум) по некоторому множеству известных $(B\bar{s}_i + \bar{e}_i)$. Задача обучения с ошибками(Learning with Errors, LWE) является обобщением задачи обучения контроля целостности (четности) данных с шумами, (Рис. 12).

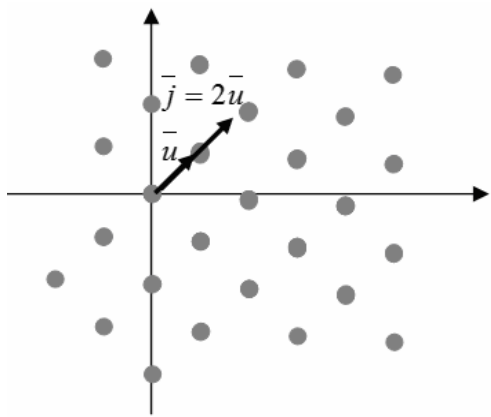 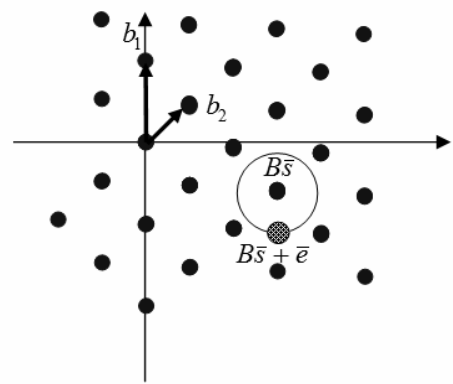

Рис. 11. Пример uSVP-задачи      Рис. 12. Пример LWE-задачи

Все криптографические системы теории решеток можно условно разделить на два типа:

- имеющие строго доказанную криптостойкость, но неэффективные по времени выполнения алгоритма шифрования/дешифрования и/или характеризующиеся быстрым ростом публичного и приватного ключей от ключевых параметров шифрования, например размерности решетки; К таким системам шифрования относятся криптографии на основе: SVP, uSVP, SIVP – задач;
- эффективные по времени шифрования/дешифрования и затратам на хранения открытого и приватного ключей, не обладающих строго доказанной криптостойкостью. К таким системам принято относить системы, основанные на некоторых частных в параметрическом смысле случаях задач теории решеток или же основанных на решетках с цикличностью образующего их базиса. К таковым, относят NTRU (Draft standard IEEE 1363.1) шифрование [9].

NTRU система шифрования основанная на задаче NTRU-свертки модулярных решеток (NTRU Convolution modular lattice, NTRU CML), которая является частным случаем CVP-задачи. Основой шифрования является операция свертки на кольце модулярных многочленов (с целыми коэффициентами). Под сверткой многочленов в данном случае понимают, их умножением, с заданным правилом свертки $x^i = 1$, где $i = const$. Например, $x^5 = 1$:

$$(2x^4 - 3x^3 + 2) \times (4x^5 + 2x - 2) \equiv -2x^4 - 6x^3 + 4x + 8.$$

Под модулярным многочленом $Z[x]/(x^n - 1) \mod k$, понимают многочлен, $P_k(x) = b_{n-1}x^{n-1} + ... + b_1 x + b_0$ коэффициенты которого являются остатком от деления, коэффициентов исходного многочлена $P(x) = b_{n-1}x^{n-1} + ... + b_1 x + b_0$ на $k$ и принадлежащие некоторому промежутку: $c_1 \leq b_i < c_2$. Обратным многочленом $P_k(x)$ по модулю $k$, является многочлена $P_k^{-1}(x)$: $P_k(x)P_k^{-1} \equiv 1(\mod k)$.

Тогда процесс шифрования будет заключаться в выборе простого $n$-показателя степени для правила свертки, малого и большого взаимно простых модулей $p, q$. Выборе многочленом $f(x), g(x) \in R$ с «малыми коэффи-

циентами». Вычисление обратных многочленов: $F_q(x) \equiv f^{-1}(x) \bmod q$ и $F_p(x) \equiv f^{-1}(x) \bmod p$. Публичным ключом является многочлен $h(x) = g(x) \cdot F_q \bmod q$. Приватным ключом многочлен $f(x)$.

Шифрование: Выберем «малый» многочлен $r(x)$. Зашифрованный текст представляет собой модулярный многочлен $m(x) \bmod p$. Зашифрованное сообщение $e \equiv p \cdot r \cdot h + m (\bmod q)$.

Расшифровка: Вычислить $a(x) \equiv e(x) \cdot f(x) (\bmod q)$, коэффициенты многочлена $A \leq a_i < A + q$. Тогда исходное сообщение $m(x) = F_p \cdot a \bmod p$.

При переходе к эквивалентному представлению многочлена в векторной форме, задаче свертки модулярного многочлена взаимно-однозначно соответствует задача свертки модулярной решетки[10]. На сегодняшний день именно эта система получила наибольшее применение среди всех систем шифрования на основе теории решеток. Причиной этого является высокая производительность алгоритма, в сочетании с малым размером публичного и приватного ключей (таб. 1). Основным недостатком NTRU шифрования является отсутствие теоретического обоснования криптостойкости, представленная в таб. 1 оценка является экспериментальной, основанной на самом быстром варианте алгоритма редукции решеток – блочном алгоритме Коркина-Золотарева(block Korkin-Zolotarev, BKZ-LLL).

Одной из наиболее исследованных систем шифрования на основе задач теории решеток является версия шифрования Айтая-Дворка(Ajtai-Dwork) лишенная ошибок расшифровки сообщения, предложенная Голдштейном, Голдвассером и Халеви (GGH) [11] - $AD_{GGH}$, основанная на uSVP-задаче. Приватный ключ, представляет собой простое число $h \in [\sqrt{K}, 2\sqrt{K}]$, где $K$ - некоторое большое число. Публичный ключ – набор из $m = O(\log K)$ чисел $0 \leq a_1, ..., a_m \leq K - 1$, которые достаточно близки к числам, кратным $\frac{K}{h}$ ( $a_n + n_{i_0} - \frac{K}{h} \leq c, c \approx 0$), плюс номер одного из этих чисел $a_{i_0}$, которое близко к нечетному числу кратному $\frac{K}{h}$. Тогда шифрование, будет заключаться в замене нулевого бита сообщения суммой случайного подмножества $\{a_i\}$, единичного бита сообщения суммой случайного подмножества $\{a_i\}+\frac{a_{i_0}}{2}$. Расшифровка будет получена в результате деления числа соответствующего биту на число $\frac{K}{h}$. Если получится мало (меньше некоторой константы) – получим нулевой бит. Много – 1 бит. Основным недостатком системы шифрования $AD_{GGH}$, является быстрый рост размеров публичного и приватных ключей в зависимости от размерности базиса решетки (Таб. 1), что затрудняет практическое применение этой системы.

Ассиметричное шифрование Реджева ($\text{Regev}_{05}$) основано на LWE-задаче со строгим доказательством криптостойкости и сочетает в себе относительно высокую скорость шифрования/дешифрования, а также сравнительно компактные публичный и приватные ключи (Таб. 1). Приватным ключом является исходный вектор $\bar{s}$, публичным $\bar{p}=B\bar{s}+\bar{e}$. Для шифрования бита $b$, выбирается некоторый вектор $\bar{e}$, шифротекст это пара: $u=B\bar{e}$ и $c=p^T\bar{e}+b\lfloor q/2 \rfloor \in Z_q^{n+1}$. Дешифровка бита заключается в вычислении $b'=c-s^T u$, исходный бит $b=0$, если $b'$ находится ближе к 0, нежели чем к $\lfloor q/2 \rfloor$, иначе $b=1$.

В июне 2009 года исследователь IBM Крейг Джентри [12] продемонстрировал реализацию гомоморфного шифрования на идеальных решетках для операций сложения и умножения ($\text{Gentry}_{mrf}$). Гомоморфное шифрование подразумевает гомоморфизм относительно некоторой операции между исходными данными и зашифрованной информацией. Такой тип шифрования позволяет реализовывать обработку зашифрованных данных (поиск по полям) без необходимости их полной расшифровки, что обеспечивает конфиденциальность информации хранимой на удаленных серверах, в том числе при обмене данными по незащищенному каналу Internet. Гипотеза о воз-можности гомоморфного шифрования была предложена Рональдом Ривестом четверть века назад, но после тщетных попыток реализации, он предположил принципиальную невозможность построения таковых систем.

Идеальная решетка – это решетка со свойствами идеала на некотором кольце чисел, т.е. результат сложения и умножения векторов в такой решетке, так же принадлежит ей самой. Это позволяет использовать свойства кольца, вместо свойств аддитивной подгруппы. В работе [12] предлагается на основе полиномиального кольца $R = Z(x)/f(x)$ и унитарных (приведенных) многочленов $f(x) \in x$, степени $\deg(f) = n$ с идеалом $J = a(x)$: $\{a(x)b(x) \bmod f(x) : b(x) \in R\}$ использовать в качестве базисов решетки коэффициенты при степенях, полученные в результате деления многочленов.

Тогда аддитивная операция примет следующий вид: $m + 2 \cdot v + j$, где $m = \{0,1\}^k$ - k битное сообщение, $v$-случайный кратчайший вектор решетки, $j \in J$ - случайный вектор из публичного ключа.

Мультипликативная операция примет вид:
$((m_1 + 2 \cdot v_1 + j_1)(m_2 + 2 \cdot v_2 + j_2) = m_1 \times m_2 + 2(m_1 v_2 + m_2 v_1 + 2v_1 v_2) + j', j' \in J$.

Параметрами шифрования будут: кольцо R; базис $B_I$ «небольшой» идеальной решетки $I$; радиусы шифрования – окружность (сфера) радиуса $r_{enc}$ в фундаментальном параллелепипеде, образованном базисом открытого ключа; радиус расшифровки – окружность (сфера) радиуса $r_{dec}$ в фундаментальном параллелепипеде образованном базисом закрытого ключа. Открытым ключом будет базис $B_{pk}$ с векторами, многократно превышающими кратчайшие, закрытым ключом будет $B_{sk}$ с векторами равными или близкими к базису «большой» идеальной решетки $J : I + J \subset R$.

Сложение определенно: $c = c_1 + c_2 \bmod B_{pk}, (m_1 + m_2) \in B_{sk}$.

Умножение: $c = c_1 \times c_2 \bmod B_{pk}, m_1 \times m_2 \in B_{sk}$, где $B_{sk}$ - фундаментальный параллелепипед, построенный на базисных векторах скрытого ключа.

Операция шифрования: $m = (m + I) \bigcap B(r_{enc}), c = m \bmod B_{pk.}$.

Операция расшифровки: $m = (c \mod B_{sk}) \mod B_I$.

Условие: $(m_1 + m_2) \in B_{sk}$ и $(m_1 \times m_2) \in B_{sk}$ - условия корректности расшифровки. Алгоритм шифрования основан на SIVP-задаче. Но для достижения достаточной криптостойкости требует повышение степени многочленов, что приводит к экспоненциальному росту операций умножения (сложность каждой для алгоритмов Карацубы и Шенхаге-Штрассена, $n^{\log_2 3}$ и $n \log_2 n \log_2 \log_2 n$ соответственно) и как результат соответствующему росту публичного и приватного ключей. Сам Джентри оценил необходимый рост производительности ЭВМ в триллион раз (через 40 лет при условии работы закона Мура), для того чтобы применение этой системы шифрования стало возможным на практике.

**Таблица 1**

Сравнение систем ассиметричного шифрования [13]

| Система шифрования | Сложность параметра | | | |
|---|---|---|---|---|
| | Криптостойкость | Размер публичного ключа | Размер приватного ключа | Шифрования, дешифрования |
| $AD_{GGH}$ | $O(n^{11}) - uSVP$ | $O(N^5 \log N)$ | $O(N^2)$ | $\approx O(n^{\log_2 c})$, c<3 |
| $\text{Regev}_{05}$ | $\widetilde{O}(n^{1.5})$-SVP | $O(N^2 \log^2 N)$ | $O(N \log N)$ | $\widetilde{O}(n)$ |
| $\text{Gentry}_{mrf}$ | $2^k$-SIVP | $\widetilde{O}(k^{3.5})$ | $\widetilde{O}(k)$ | $\widetilde{O}(k^7)$ |
| NTRU | $10^{0.0826n-2.58}$ сек., | $\approx \frac{1}{2} N \log_2 \frac{N}{4}$, | $\approx \frac{N(n-k) \log_2 \frac{N}{4}}{2n \log_p q}$, | $O(n^2)$ |

, где $n$ - размерность решетки, N-объем базиса решетки в битах, $k$ - уровень защищенности в битах относительно симметричного алгоритма шифрования, $p$ -малый модуль, $q$ - большой модуль.

Таким образом, необходимость устранения существующего противоречия между «криптостойкими теоретическими» и «эффективными эмпирическими» системами ассиметричного шифрования на основе задач теории решеток, открывает перспективное направление фундаментальных и прикладных математических исследований в области постквантовой криптографии.